\documentstyle[psbox]{WORKSHOP}

\begin{document}

\title{\bf{Different Types of X-ray Variability Observed
in Micro-Quasar GRS 1915+105 with the IXAE}}

\author{
S. Naik$^1$, P. C. Agrawal$^1$, B. Paul$^1$, A.R.Rao$^1$, J.S.Yadav$^1$ and
S. Seetha$^2$
\\[12pt]  % TO BE SPACED WITH ONE LINE
$^1$ Tata Institute of Fundamental Research, Homi Bhabha Road, Mumbai, 400 005, India \\
$^2$ ISRO  Satellite Center, Airport Road, Vimanapura P.O., Bangalore - 560017,
India \\
{\it E-mail (S. Naik): sachi@tifr.res.in}
}

\abst{
The galactic micro-quasar source GRS 1915+105 has been observed with
the Indian X-ray Astronomy Experiment (IXAE) several times over a period
extending from 1996 to 2000. The source  has been observed in different
luminosity states and exhibits a rich variety of random and regular X-ray
variability over a wide range of time scales. It also shows a unique
variability characteristic namely the occurrence of quasi-regular X-ray
bursts with period ranging from 20 s to about 100 s which have slow rise
and fast decay. The X-ray spectrum of these bursts is hardest near the end
of decay. In the low-hard state and bright-soft state with rapid
variability, the source exhibits quasi-periodic oscillations (QPOs) with
variable frequency which is luminosity dependent. Regular X-ray intensity
dips have been detected coincident with the occurrence of a giant radio
outburst. The source made a transition from a low-state to a high-state in
1999 in less than a day and from a soft-bright state to a hard-low state in 
2000 in less than 100 minutes during the IXAE observations. Details of these 
and other variability characteristics of the source are presented in this 
paper.}

\kword{ accretion: accretion disks - black hole physics - X-rays: stars - stars: individual - GRS 1915+105}

\maketitle
\thispagestyle{empty}

\section{INTRODUCTION}

The galactic X-ray source GRS~1915$+$105 discovered in 1992 shows superluminal
motion and other radio characteristics similar to those of quasars and hence
has been termed as a `micro-quasar'. From resemblance of its spectral and 
temporal properties with that of superluminal Galactic radio source GRO 
J1655$-$40 
(Zhang et al. 1997), GRS 1915$+$105 is believed to be a black hole. The source 
is very bright in X-ray, emitting at a luminosity of more than 10$^{39}$
erg s$^{-1}$. It exhibits strong variability over a wide range of 
time scales in X-ray, infrared and radio bands. The X-ray emission is 
characterized by chaotic variability as well as narrow quasi-periodic 
oscillations (QPOs) at centroid frequency  in the range of 0.001 $-$ 67 Hz. 
It is found that the intensity dependent narrow QPOs are a characteristic 
property of the hard-state which is absent in the soft-state. Based on 
extensive X-ray studies the behaviour of the source can be classified in  
two distinct states: the spectrally hard-state, dominated by a power-law 
component when the QPOs are present and the soft-state, dominated by thermal 
emission when the QPOs are absent.

The micro-quasar GRS 1915$+$105 shows a variety of features
in the X-ray light curve. Large, eclipse-like dips in the
X-ray flux, called as brightness sputters, with a recurrence time
of about 250 s, extremely large amplitude oscillations with an
amplitude of $\sim$ 3 Crab and recurrence periods of 30-100 s,
fast oscillations in the X-ray flux, brief flares (lasting for
a few seconds) following the low level X-ray flux (lull) and
several complex type of regular and quasi-regular variations
in the X-ray flux are observed in the light curves at different 
times. Wide range of transient activity, including the regular and
quasi-regular bursts with a secondary (and tertiary) weaker
burst following the primary burst, are described as a result
of thermal instability in the inner accretion disk producing
short duration luminosity fluctuations which are observed as
bursts. Belloni et al. (2000) have made an extensive study on
the X-ray emission of the source and classified all the publicly
available RXTE/PCA observations from 1996 January to 1997 December
into 12 different classes on the basis of structure of the X-ray
light curve and the nature of the color diagram. From the above study,
they found that the source variability is restricted into
three basic states, a hard-state with non-observability of the
inner accretion disk and two soft-states with visible inner accretion
disk and at different temperatures depending on the different values of
mass accretion rate.

\begin{table*}[t]
\caption{X-ray Observations of GRS 1915$+$105 with the IXAE}
\vskip 0.4cm
%\begin{center}
\begin{tabular}{lllllll}
\hline
\hline
Year  &\multicolumn {2}{c}{Observation} &Useful &Main X-ray variability Characteristics\\
of Obs. &start  &End &time(s) & &\\
\hline
\hline
1996    &July 20&July 29&8,850     &Strong erratic intensity variations on time scale of \\
        &       &       &          &0.1$-$10 s and QPOs in a frequency range of 0.62$-$0.82 Hz\\
        &       &       &          &are observed (Paul et al. 1997). The intensity variations \\
        &       &       &          &are described as sum of shots in the light curve (Paul et \\
	&	&	&	   &al. 1998a).\\
\\
1997 &June 12   &June 29   &39,300 &Different types of intense, quasi-regular X-ray bursts are\\
     &       &       &             &observed with slow-rise and fast decay and interpreted as \\
     &       &       &             &the disappearance of matter through event horizon of the \\
     &	     &	     &             &black hole (Paul et al. 1998b).\\
\\
1997 &August 07 &August 10 &20,700 &Classification of quasi-regular and irregular X-ray bursts\\
     &       &       &             &into different classes and explained as the appearance and \\
     &       &       &             &disappearance of the advective disk over its viscous timescale \\
     &       &       &             &(Yadav et al. 1999).\\
\\
1999 &June 06 &June 17     &37,740 &Transition from a low-hard state to high-soft state is observed.\\
     &	      &            &       &Detection of a series of soft X-ray dips which are explained \\
     &       &       &             &as the cause for mass ejection due to the evacuation of \\      
     &       &       &             &matter from an accretion disk producing a huge radio jet in \\
     &       &       &             &GRS 1915+105 (Naik et al. 2001a).\\
\\
2000 &June 18   &June 27  &29, 460 &Transition from a high-state with regular and periodic \\
     &       &       &             &bursts with slow rise and sharp decay (class $\rho$) to a \\
     &       &       &             &low-hard state is observed within 1.5 hrs (Naik et al. 2001b).\\
\hline
\hline
\end{tabular}
%\end{center}
\end{table*}

Attempts have been made to explain the observed X-ray properties of the
source with the properties at longer wavelengths. Simultaneous observations
at shorter and longer wavelengths have established the correlation between the
X-ray variability and the emission at radio and infrared wavelengths
(Fender et al. 1999 and reference therein). They showed that the emission
at longer wavelengths is associated with the plasma ejection from the
instabilities in the inner accretion disk. Naik \& Rao (2000) examined
the radio properties of the source during all the 12 different X-ray classes.
They showed that the huge radio flares are associated with the soft-dips
in X-ray light curve (classes $\theta$ and $\beta$) where as the higher flux
density at radio wavelengths with flat spectrum is associated with the X-ray
low-hard states referred as ``plateau'' state (classes $\chi1$ and $\chi3$).
It is found that in all the other X-ray classes of Belloni et al. (2000)
classification, the source remains in a radio-quiet state.

In this paper, we present a summary of the main variability characteristics
of the source observed in different states with the Pointed mode
Proportional Counters (PPCs) of IXAE. It was
found to be in different brightness states during the five different epochs
of observations. 

\section{OBSERVATIONS}
The X-ray observations of GRS 1915$+$105 were carried out from 1996 
July to 2000 June at five different epochs with the PPCs of
the IXAE on board the Indian satellite IRS-P3. The IXAE includes three
co-aligned and identical, multi-wire, multi-layer proportional counters,
filled with a gas mixture of 90\% Argon and 10\% Methane at a pressure of
800 torr, with a total effective area of 1200 cm$^{2}$, covering 2 to 18 keV
energy range with an average detection efficiency of about 60\% at 6 keV.
For a detailed description of the PPCs refer to Agrawal (1998). A log of
the observations of the source GRS 1915$+$105 from 1996 July to 2000 June
is given in the Table. 1 with the useful period of observations at each epoch
and important results obtained  in each observation.

\begin{figure}
\vskip 7.5 cm
\includegraphics{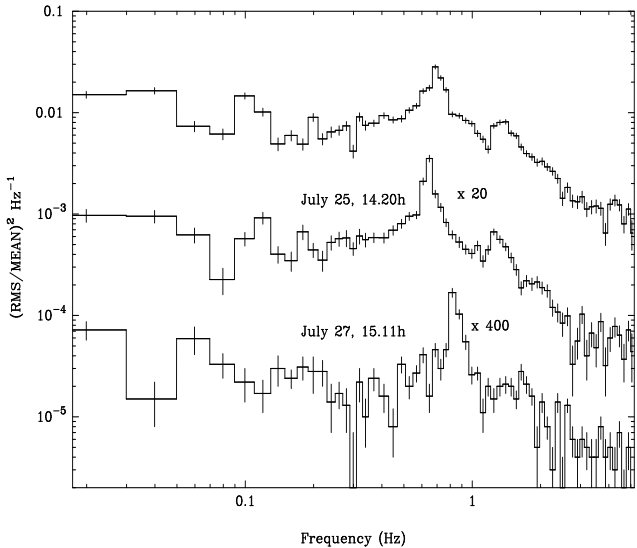}
\caption{The Power density spectra of the X-ray source GRS1915+105 observed 
with IXAE in 2$-$18 keV energy band. A strong peak at a frequency of 0.7 Hz 
is clearly visible. The two top plots show power density spectra with peaks at frequencies 
0.62 Hz and 0.82 Hz respectively indicating variation of the QPO peak 
frequency. The dates of observations are given in the figure. The first 
harmonic at a frequency of twice that of the main peak is seen in all 
the three plots.}
\end{figure}

\begin{figure}[t]
\vskip 7.5 cm
\includegraphics{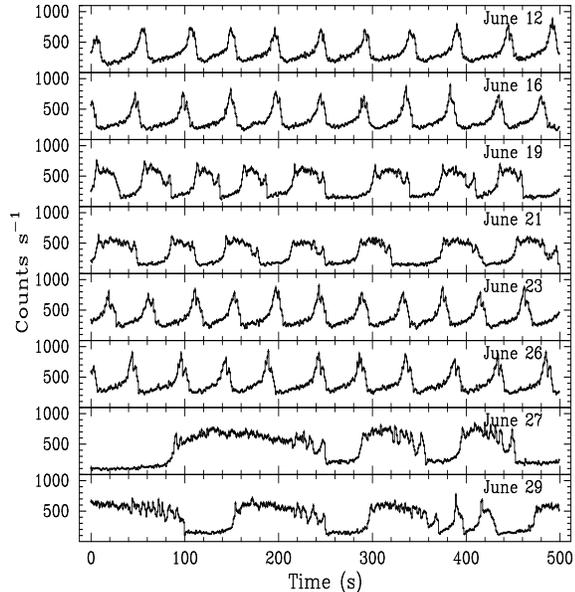}
\caption{The regular (first, second, fifth, and sixth panels from the top), 
irregular (third and fourth panels), and long (seventh and eighth panels) 
bursts observed in GRS 1915+105 with one of the PPCs in June 1997 as 
indicated in the respective panels. }
\end{figure}

\begin{figure}[t]
\vskip 5.6 cm
\includegraphics{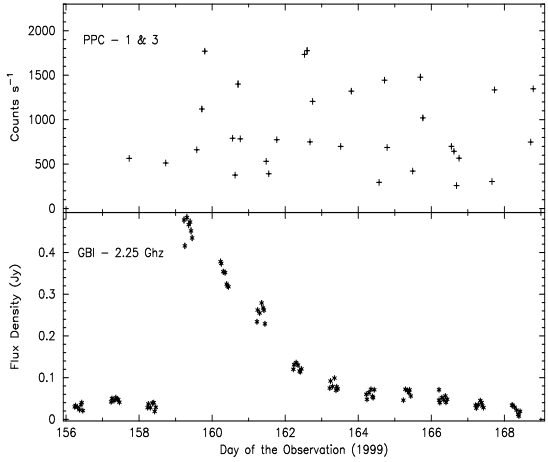}
\caption{The X-ray light curve for GRS 1915$+$105 with the PPCs (averaged 
over each orbit) in the energy range 2$-$18 keV and radio flux at 2.25 GHz 
with NSF-NRAO-NASA Green Bank Interferometer. The transition of the source 
from a low-hard state to a high-state was observed within a day as shown in 
top panel of the figure.}
\end{figure}

\begin{figure}[t]
\vskip 6.8 cm
\includegraphics{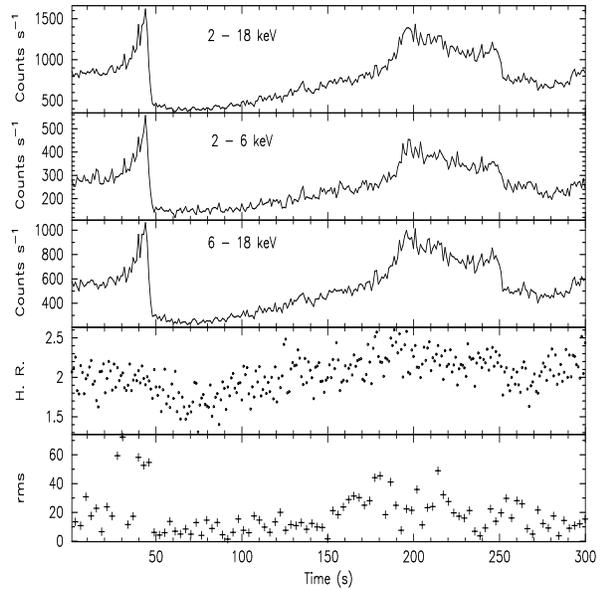}
\caption{Detailed light curves of GRS 1915+105 during an X-ray dip observed 
with the PPCs in 1999 June, with 1 s bin size in the energy range of 2-18 keV, 
2-6 keV, and 6-18 keV. The dip is identical in shape in all the energy ranges. 
The hardness ratio (HR) is plotted in the fourth panel of the figure, which 
indicates the soft nature of the spectrum during the dips. The lower panel 
shows the variation in the rms of the source during the dip and non-dip 
regions.}
\end{figure}

\begin{figure}[t]
\vskip 10.9 cm
\includegraphics{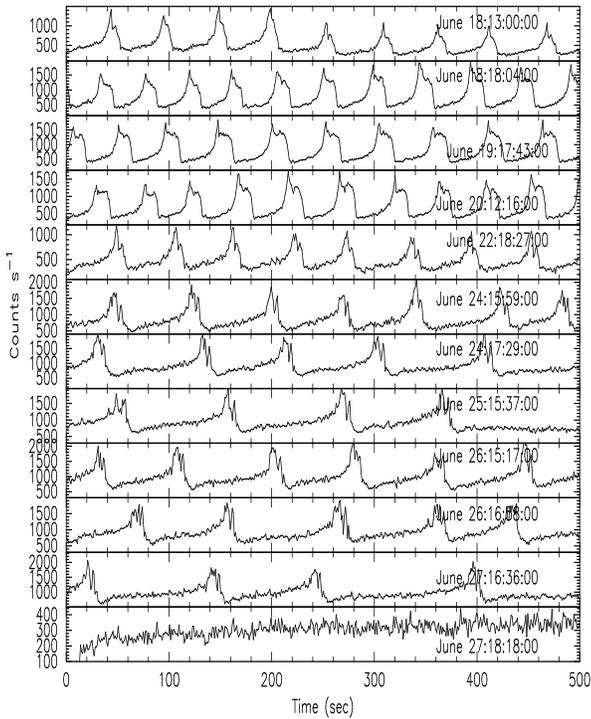}
\caption{ The X-ray light curves for GRS 1915$+$105 with the PPCs in the energy
range 2 $-$ 18 keV in 1 s time integration mode during the IXAE observation in 
2000 June. Change in burst duration from 2000 June 18 to 27 is shown. The 
alphabets represent the data from different orbits of PPC observations.}
\end{figure}

\begin{figure}[t]
\vskip 10.9 cm
\includegraphics{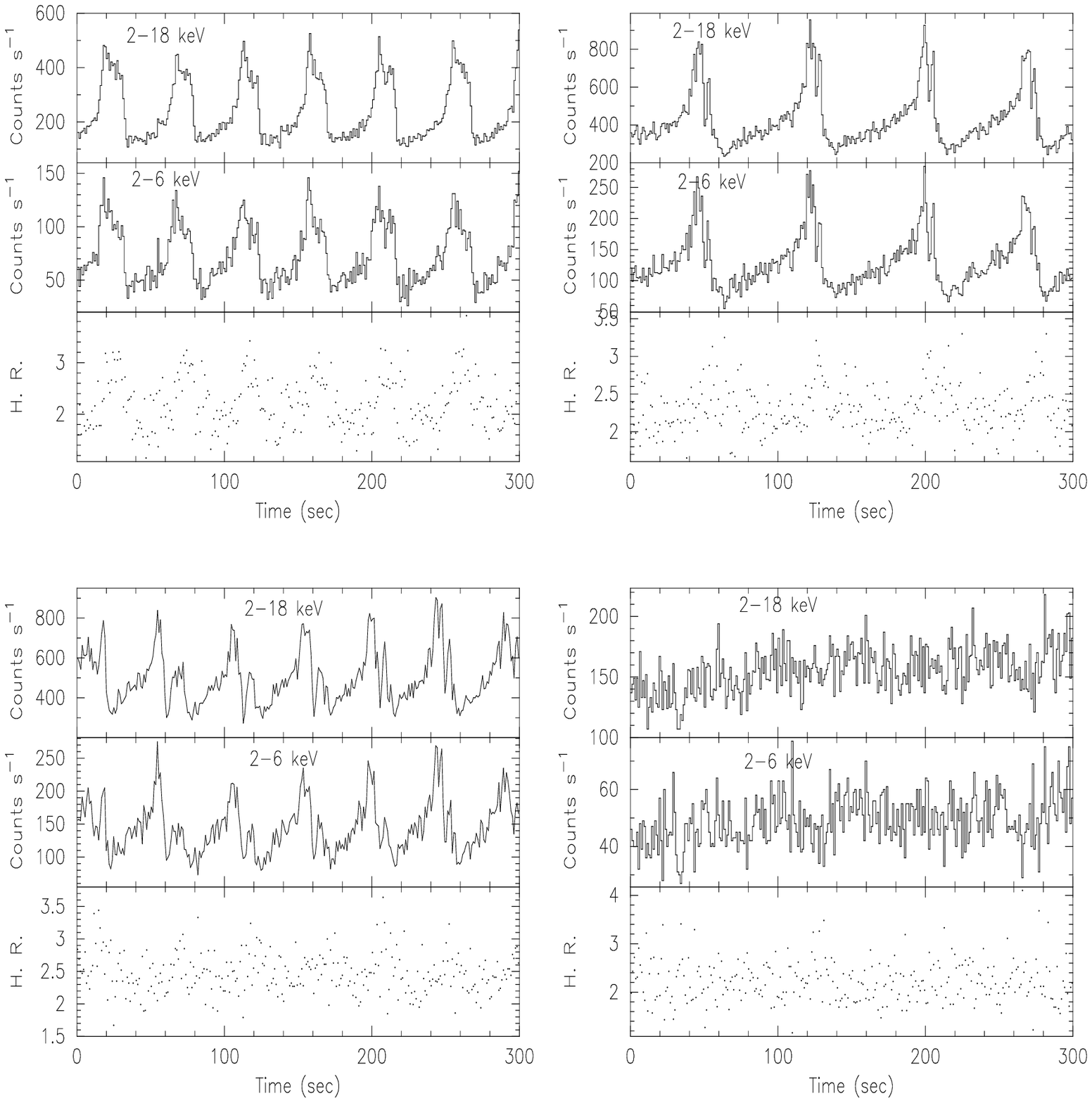}
\caption{ The X-ray light curves of the source GRS 1915$+$105 obtained from IXAE
during four different intervals of 2000 June observation are shown in the
energy ranges 2$-$18 keV and 2$-$6 keV with hardness ratio (H. R. = Count rate
in 6 $-$ 18 keV range/count rate in 2$-$6 keV range). The appearance of
secondary (and tertiary peaks) in the later light curves can be seen.}
\end{figure}

\section{RESULTS}
Following are the main results obtained from the IXAE observations of GRS 1915$+$105 from 1996 to 2000. 

(i) Intense and erratic intensity variations, including sub-second flares, on 
time scales of 0.1 s $-$ 1 s were observed in the energy range of 2 $-$ 18 keV 
during 1996 July observations of the source. Strong and narrow QPOs in a 
frequency range of 0.62 $-$ 0.82 Hz with a rms fraction of about 9\% were also 
detected as shown in Figures 1. Besides the low frequency QPOs (0.62 $-$ 0.82 
Hz), a less prominent peak at $\sim$ 1.4 Hz which is the first harmonic
of the main peak was also detected with the PPCs. The erratic intensity 
variations on sub-second time scale, low frequency QPOs and super-Eddington 
luminosity, which are also observed in other stellar black hole X-ray binaries, support the hypothesis that the compact object is most likely a black hole 
(Paul et al. 1997). Time lag of of 0.2 to 0.4 s of hard X-ray (6$-$18 keV) 
photons relative to that of the soft (2$-$6 keV) photons was also detected 
which supports the idea of the generation of hard X-ray photons by Compton 
up-scattering from the high energy clouds near the source of soft X-rays. 
The observed intensity variations in the light curves are described by a shot 
noise model (Paul et al. 1998a). Chaotic variability in the intensity is 
explained as resulting from randomly occurring shots with exponential rise 
and/or decay.

(ii) Different kinds of periodic and quasi-periodic bursts are 
observed with the PPCs during different epochs. These bursts are classified 
into four different classes on the basis of the different time scales such as
(1) regular bursts having a slow rise and fast decay lasting for $\sim$ 15 s 
and recurring every 21 s; (2) regular bursts, having a slow rise and sharp 
decay lasting for $\sim$ 20 s and recurring every 46 s; (3) quasi-regular 
bursts of variable duration with slow rise and sharp decay; and (4) irregular 
bursts, with duration of a few tens to a few hundreds of seconds, followed by 
sharp decay. Figure 2 shows the structures of different types of regular 
and quasi-regular bursts, which were detected from the 1997 June observations 
of the source. These bursts are characterized by an exponential rise with a 
timescale of about 7$-$10 s and a sharp linear decay in 2$-$3 s. The slow 
rise time of the bursts is explained as the infall time of matter from the 
accretion disk and the fast decay is due to the disappearance of matter 
through the event horizon of the black hole (Paul et al. 1998b). The 
repetition of these regular bursts can be due to the material influx into 
the inner disk because of the oscillations of the shock front far away from 
the compact object. The transition from the quiescent state to the burst 
state and vice versa is 
explained by the appearance and disappearance of the advective disk over 
its viscous time scale.

(iii) A series of X-ray soft-dips of durations in the range of 20$-$160 s were 
detected in the X-ray light curves during the transition of the source from a 
low-hard state to high-state in 1999 June observations. The transition takes 
place within a few hours as indicated by the RXTE/ASM and IXAE light curves 
(Naik et al. 2001). The lower panel in Figure. 3 shows a huge radio flare 
at 2.25 GHz which was observed with the NSF-NRAO-NASA Green Bank Interferometer 
following the change of state in X-rays as observed with the PPCs (upper 
panel). Detailed analysis showed that this huge radio flare was associated
with the occurance of a series of X-ray soft dips as shown in Figure 4. 
During the the dips, the low frequency QPOs were absent and the spectrum 
was soft with less variability compared to the non-dip regions. 
In contrast to this, during the non-dip regions, the light curve of the 
source is dominated by a pronounced variability with the presence of 
low-frequency ($\sim$ 4 Hz) QPOs. The spectrum of the source is also hard 
during the non-dip regions. These X-ray dips are suggested as the cause of 
mass ejection due to evacuation of matter from the accretion disk producing 
a huge radio jet.

(iv) A change of state of the source from a high-state (consisting of regular 
and periodic bursts) to a low-state was observed during the 2000 June 
observations. At the beginning of the observations, bursts with recurrence 
time of $\sim$ 40 s were detected which changed to $\sim$ 120 s towards the 
end of the observations. These regular and periodic bursts disappeared from 
the X-ray light curve within about 1.5 hours (one orbit of the satellite 
IRS-P3) and the source changed to a low-hard state as shown in the last 
panel of Figure 5. The spectrum of the source becomes harder as the burst 
proceeds and becomes hardest at the end of the decay. The bursts are marked 
by the presence of a secondary (tertiary) peak during their decay phase
as seen in Figure 6. The secondary (tertiary) peak resolves gradually from 
the primary peak with the increase in the burst recurrence time and finally 
all the bursts disappear from the light curve when the state of the source 
is changed as shown in the last panel of Figure 6. 

\section{DISCUSSION}
Though the X-ray states of the microquasar GRS 1915$+$105 are classified
into the canonical low-hard state and high-soft states as seen in other
stellar black hole sources, it does exhibit extended low-hard states 
(Morgan et al. 1997).
After the extended low-hard state, the source switches into a high-soft
state which can be divided into different types, such as bright state,
chaotic state and flaring state (Rao et al. 2000). The low-hard states
are characterized by low frequency QPOs at $\sim$ 2 $-$ 3 Hz which according
to the model of Molteni et al. (1996) is due to the oscillations
of the shock. The QPO frequency varies inversely with the shock radius. 
Muno et al. (1999) found that the QPO frequency 
increases with the disk temperature and decreases with the disk radius. The 
regular and periodic bursts seen in the X-ray light curves are explained by the 
periodic infall of matter onto the black hole from an oscillating shock front 
(Molteni et al. 1996). According to this model, a shock is produced in 
the sub-Keplerian component of the accretion disk because of the centrifugal 
barrier. The shock starts oscillating if the cooling time of the post shock  
halo matches with the material infall time. The oscillation period depends on 
the angular momentum and the viscosity of the accreted matter from the 
companion star. If the accreted matter has high angular momentum and low 
viscosity, then the oscillation period is higher. They tried to explain all 
the observed properties of the bursts by the above model. The bursts result 
from a catastrophic infall of matter, piled up behind the shock, onto 
the black hole which increases the temperature and hence the X-ray intensity.
Paul et al. (1998b) and Yadav et al. (1999) have made a detailed study
of such intensity variations using the data obtained with the IXAE.
Yadav et al. (1999) concluded that the repeated intensity variation cannot 
be attributed to inner disk evacuation. This is due to the viscous time scale 
arguments as well due to the fact that the two intensity states are quite similar 
to the low-hard and high-soft state of the source. They invoked the 
two-component accretion-flow model (TCAF) of Chakrabarti \& Titarchuk (1995) 
to conclude that the rapid changes are due to the appearance and disappearance 
of advective disk covering the standard thin disk without any requirements of 
mass ejection. 

\section{Acknowledgments }

We acknowledge the contributions of the scientific and technical staff
of TIFR, ISAC and ISTRAC for the successful fabrication, launch and
operation of the IXAE. It is a pleasure to acknowledge constant  support
of Dr. K. Kasturirangan, Chairman ISRO, Shri K. Thyagarajan, Project Director 
IRS$-$P3 satellite, Shri J. D. Rao and his team at ISTRAC, Shri P. S. Goel, 
Director ISAC and the Director of the ISTRAC. We thank the NSF-NRAO-NASA
Green Bank Interferometer group for making the data publicly available.

\section*{References}
Agrawal, P. C., {\it Perspective in High Energy Astronomy and Astrophysics;
Proceedings of the International Colloquium}, Aug. 12-17, 1996 at TIFR,
Mumbai, eds: P. C. Agrawal and P. R. Viswanath, University Press, Hyderabad,
India, p 408\\
Belloni, T., Klein-Wolt, M., Mendez, M., van der Klis, M., and van Paradijs, J.  2000, A\&A, 355, 271 \\
Chakrabarti, S. K., \& Titarchuk, L. G. 1995, ApJ, 455, 623\\
Fender, R. P., Garrington, S. T., McKay, D. J., Muxlow, T. W. B., Pooley, G. G., Spencer, R. E., Stirling, A. M., Waltman, E. B.  1999, MNRAS, 304, 865 \\
Molteni, D., Sponholz, H., \& Chakrabarti, S. K. 1996, ApJ, 457, 805 \\
Morgan, E. H., Remillard, R. A., \& Greiner, J. 1997, ApJ, 482, 993\\
Muno, M. P., Morgan, E. H., \&  Remillard, R. A. 1999, ApJ, 527, 321 \\
Naik, S., \& Rao, A. R. 2000, A\&A, 362, 691\\
Naik, S., Agrawal, P. C., Rao, A. R. et al. 2001a, ApJ, 546, 1075\\
Naik, S., Agrawal, P. C., Rao, A. R. et al. 2001b, in preparation \\
Paul, B., Agrawal, P. C., Rao, A. R., et al. 1997, A\&A, 320, L37\\
Paul, B., Agrawal, P. C., Rao, A. R., et al. 1998a, A\&AS, 128, 145\\
Paul, B., Agrawal, P. C., Rao, A. R., et al. 1998b, ApJ, 429, L63\\
Rao, A. R., Yadav, J. S., \& Paul, B. 2000, ApJ, 544, 443 \\
Yadav, J. S., Rao, A. R., Agrawal, P. C., et al. 1999, ApJ, 517, 935\\
Zhang, S. N., Cui, W., \& Chen, W. 1997, ApJ, 482, L155\\

\end{document}